\documentclass[a4paper]{article}

\usepackage{INTERSPEECH2019}
\usepackage{amsmath}
\usepackage{booktabs}
\usepackage{multirow}

\title{Connecting and Comparing Language Model Interpolation Techniques}
\name{Ernest Pusateri, Christophe Van Gysel, Rami Botros\sthanks{\hspace{0.5\baselineskip}No longer at Apple Inc.}, Sameer Badaskar, Mirko Hannemann,\\Youssef Oualil, Ilya Oparin}
\address{
  Apple Inc., USA
}
\email{\{epusateri,cvangysel,badaskar,mirko\_hannemann,youalil,ioparin\}@apple.com}

\begin{document}

\maketitle

\begin{abstract}

In this work, we uncover a theoretical connection between two language model interpolation techniques, count merging and Bayesian interpolation. We compare these techniques as well as linear interpolation in three scenarios with abundant training data per component model. Consistent with prior work, we show that both count merging and Bayesian interpolation outperform linear interpolation.  We include the first (to our knowledge) published comparison of count merging and Bayesian interpolation, showing that the two techniques perform similarly.  Finally, we argue that other considerations will make Bayesian interpolation the preferred approach in most circumstances.

\end{abstract}
\noindent\textbf{Index Terms}: speech recognition, language modeling, language model interpolation, count merging, Bayesian interpolation

\section{Introduction}
\label{sec:introduction}
Virtual assistants such as Apple's Siri continue to gain in popularity.  Their appeal comes in part from their versatility.  The most advanced virtual assistants can respond to requests in a broad range of domains, from simple commands like ``set my alarm'' to complex questions about rare named entities.  This requirement to handle requests from a wide variety of domains poses a particular challenge when building the language model (LM) for the speech recognition component of such a system.

Neural network LMs have been shown in recent years to outperform $n$-gram models on a variety of tasks (e.g. \cite{Chelba2013onebillionwords}), and so one might consider their application here.  However, $n$-gram models maintain advantages in decoding efficiency and in their ability to be quickly trained with very large amounts of data.  These advantages often make an $n$-gram model the preferred choice for the first recognition pass of modern real-world systems.  Further, even when a neural network LM is used, it is often interpolated with an $n$-gram model.  For these reasons, this work will assume the use of $n$-gram models.

To build an $n$-gram model that covers a broad range of domains, a common approach is to build separate domain-specific LMs and then apply an interpolation technique to combine them in a way that's optimized for the target use case.  Linear interpolation is one of the simplest and most commonly used of these techniques \cite{JelinekMercer80}.  Count merging \cite{Bacchiani,Hsu}  and Bayesian interpolation \cite{Allauzen} are more sophisticated techniques that use word history statistics to achieve better performance, especially in cases where component domains vary widely in content or quantity of training data.  Despite their high-level similarity, work on count merging and Bayesian interpolation has proceeded in separate threads in the research literature.  Accordingly, to our knowledge, there is no published comparison of these two techniques.  In this work we uncover a theoretical connection between count merging and Bayesian interpolation.  We then compare the performance of linear interpolation, count merging and Bayesian interpolation on three large data sets.

\section{Interpolation Techniques}
\label{sec:interpolation}
We start by formulating LM combination as history-dependent linear interpolation.  Assume the target use case covers multiple domains.  Then the probability for a word, $w$, given a word history, $h$ can be expressed as the following, where $i$ represents a domain:
\begin{align}
\label{eq:p_w_h}
p(w|h) & = \sum_{i}p(w,i|h)\text{,} \\
       & = \sum_{i}p(i|h)p(w|i,h)\text{.} \label{eq:p_w_h2}
\end{align}
In theory, we could learn a parameter corresponding to $p(i|h)$ for every $i$ and $h$.  But this is rarely feasible in practice.  Instead, we will derive an estimate of $p(i|h)$ that requires a small number of learned parameters.  To do that, we first apply Bayes' rule to $p(i|h)$:
\begin{equation}
\label{eq:p_i_h}
  p(i|h) = \frac{p(i)p(h|i)}{\sum_{j}p(j)p(h|j)}\text{.}
\end{equation}
Let $\lambda_i$ be a learned parameter corresponding to $p(i)$ and $p^{\text{Interp}}(h|i)$ be an estimate of $p(h|i)$:
\begin{equation}
  \label{eq:p_interp}
  p^{\text{Interp}}(i|h) = \frac{\lambda_ip^{\text{Interp}}(h|i)}{\sum_{j}\lambda_jp^{\text{Interp}}(h|j)}\text{.}
\end{equation}
Now, assume we build a component LM for each domain.  Let $p_i(w|h)$ be the probability given to word $w$ in the context of $h$ by the $i$th component LM.  Plugging this and our estimate for $p(i|h)$ into Equation \ref{eq:p_w_h2}:
\begin{equation}
  \label{eq:p_interp_w_h}
  p^{\text{Interp}}(w|h) = \sum_i \frac{\lambda_ip^{\text{Interp}}(h|i)}{\sum_{j}\lambda_jp^{\text{Interp}}(h|j)}p_i(w|h)\text{.}
\end{equation}
We now have a formulation that assumes the existence of estimates for per-component history probabilities, $p^{\text{Interp}}(h|i)$, and has a small number of learned parameters, $\lambda_i$.  Next, we show that linear interpolation, count merging and Bayesian interpolation differ only in how they compute $p^{\text{Interp}}(h|i)$.

\subsection{Linear Interpolation}
Linear interpolation makes the strong assumption that $p(h|i)$ is independent of $i$.\footnote{This is more commonly but equivalently stated as assuming $p(i|h)$ is independent of $h$.}  Thus, Equation \ref{eq:p_interp_w_h} reduces to:
\begin{equation}
  \label{eq:p_l_i}
  p^{\text{LI}}(w|h) = \sum_{i}\lambda_ip_i(w|h)\text{.}
\end{equation}

\subsection{Count Merging}
Count merging uses the maximum likelihood estimate for $p^{\text{Interp}}(h|i)$.  It is computed as follows, where $N_i$ is the total number of words in corpus $i$ and $c_i(h)$ is the count of history $h$ in corpus $i$.
\begin{equation}
\label{eq:p_cm_h_i}
p^{\text{CM}}(h|i) = \frac{c_i(h)}{N_i}\text{.}
\end{equation}
Plugging this into Equation \ref{eq:p_interp_w_h}:
\begin{equation}
\label{eq:p_cm_w_h}
p^{\text{CM}}(w|h) = \sum_i\frac{\lambda_i\frac{c_i(h)}{N_i}p_i(w|h)}{\sum_{j}\lambda_j\frac{c_j(h)}{N_j}}\text{.}
\end{equation}
This is not the conventional formulation of count merging, as presented in \cite{Bacchiani} and \cite{Hsu}:
\begin{equation}
\label{eq:p_cm_w_h2}
p^{\text{CM}'}(w|h) = \sum_i\frac{\beta_ic_i(h)p_i(w|h)}{\sum_j\beta_jc_j(h)}\text{.}
\end{equation}
However, the two formulations are equivalent in an important way.  Define $\beta_i$ as follows, where K is a constant whose value does not affect $p^{\text{CM}'}(w|h)$:
\begin{equation}
\label{eq:beta_i}
  \beta_i = \lambda_i\frac{K}{N_i}\text{.}
\end{equation}
%
We see that for any set of $\lambda_i$ in Equation \ref{eq:p_cm_w_h}, there is a set of $\beta_i$ in Equation \ref{eq:p_cm_w_h2} that will make $p^{\text{CM}}(w|h)$ equal to $p^{\text{CM}'}(w|h)$, and vice versa. 

Note that the formulation in Equation \ref{eq:p_cm_w_h} has a few advantages over the one in Equation \ref{eq:p_cm_w_h2}.  First, it makes clear that it is the history counts relative to the size of their respective corpora that matter rather than the absolute counts.  Second, the $\lambda_i$ are more constrained than the $\beta_i$, which may make optimization easier.  Finally, unlike the $\beta_i$ in Equation \ref{eq:p_cm_w_h2}, we can interpret the $\lambda_i$ in Equation \ref{eq:p_cm_w_h} as component priors.  In addition to making interpretation of learned $\lambda_i$ easier, this allows for the $\lambda_i$ to be set using prior knowledge (e.g. an educated guess about domain usage frequencies), something that would be difficult to do with the $\beta_i$ parameters.

Note also that given a learned set of $\beta_i$, we can easily compute the implied values for $\lambda_i$.  Solving Equation \ref{eq:beta_i} for $\lambda_i$ and taking into account that $\lambda_i$ sum to $1$:
\begin{equation}
  \lambda_i = \frac{\beta_iN_i}{\sum_j\beta_jN_j}\text{.}
\end{equation}

\subsection{Bayesian Interpolation}
While count merging uses a maximum-likelihood estimate for $p(h|i)$ (Equation \ref{eq:p_cm_h_i}), Bayesian interpolation uses something more robust.  Specifically, $p(h|i)$ is computed as the probability of word sequence $h$ given by the $i$th component LM, which we denote $p_i(h)$\footnote{For example, in the case where component model $i$ is a $4$-gram model, and $h_{-1}$, $h_{-2}$, and $h_{-3}$ represent the words comprising the history $h$, $p_i(h)$ would be the product of the unigram probability of $h_{-3}$, the bigram probability of $h_{-3}h_{-2}$ and the trigram probability of $h_{-3}h_{-2}h_{-1}$.}.
\begin{equation}
  \label{eq:p_bi_h_i}
  p^{\text{BI}}(h|i) = p_i(h)\text{.}
\end{equation}
Plugging this into Equation \ref{eq:p_interp_w_h}, we see that $p(w|h)$ is computed as follows:
\begin{equation}
\label{eq:p_bi_w_h}
  p^{\text{BI}}(w|h) = \sum_i\frac{\lambda_ip_i(h)p_i(w|h)}{\sum_j\lambda_jp_j(h)}\text{.}
\end{equation}
Note that \cite{Bacchiani} applied count merging using ``expected counts'' for word histories.  Expected counts were computed by multiplying the total number of words in the training corpora by an estimate of $p(h,i)$ where $p(h,i)$ was estimated ``in a manner which may reserve probability mass for unobserved events.''  Below we show one way of computing expected counts that is consistent with that description, where the first term is the total number of words in the training corpora and the second is an estimate of $p(h,i)$.
\begin{equation}
  \label{eq:c_e_i}
  c^{\text{expected}}_i(h) = \left(\sum_jN_j\right)\left(\frac{N_i}{\sum_jN_j}p_i(h)\right)\text{.}
\end{equation}
Substituting Equation \ref{eq:c_e_i} into Equation \ref{eq:p_cm_w_h} results in Equation \ref{eq:p_bi_w_h}.  Thus, count merging with this definition of expected counts is equivalent to Bayesian interpolation.

\section{Static Interpolation}
\label{sec:static}
Whatever interpolation technique we use, we'd like to end up with a single $n$-gram LM.  However, the backoff structure of $n$-gram LMs presents a complication.  Exact computation of the interpolated model probabilities for contexts unseen in any component LM requires that we maintain separate component LMs and perform interpolation online.  So, we apply a commonly used approximation to create a statically interpolated model.  First we fix the set of $n$-grams in the statically interpolated model to be the union of all the $n$-grams seen in the component models.  Then, we interpolate separately at each $n$-gram level.  Finally, we recompute the backoff weights so that word probabilities for all histories add up to 1.

\section{Related Work}
\label{sec:related}
We are largely concerned with connecting two threads of research, one on count merging and one on Bayesian interpolation.  Count merging appears to be first formally described in \cite{Bacchiani2003} (and elaborated on in \cite{Bacchiani}), while widely known (in some form) before that.  \cite{Bacchiani2003} and \cite{Bacchiani} show that count merging using two data sources is a special case of maximum a posterior (MAP) adaptation.  \cite{Hsu} starts from the formal definition of count merging in \cite{Bacchiani} and describes a more general framework where history count is one of a handful of features used to compute history-dependent weights.

Bayesian interpolation is first described in \cite{Weintraub} for two data sources and then extended in \cite{Allauzen}.  The presentation in \cite{Allauzen} differs from ours in a couple of ways.  First, their formulation contains an extra layer of interpolation parameters.  Second, they assume estimates for the parameters analogous to $\lambda_i$ are available as inputs to the interpolation process, while we learn those parameters on a validation set.

\cite{Liu} represents another vein of related work.  We note in Section \ref{sec:interpolation} that while it is theoretically possible to learn parameters corresponding to $p(i|h)$ in Equation \ref{eq:p_w_h}, it is rarely practical.  \cite{Liu} starts from that same observation but does not make the moves in Equations \ref{eq:p_i_h}, \ref{eq:p_interp} and \ref{eq:p_interp_w_h}.  Instead they explore robust ways of more directly estimating parameters for $p(i|h)$, including various forms of MAP adaptation.

\section{Experiments and Results}
\label{sec:experiments}
\begin{table}[th]
  \caption{Results on the Billion Word scenario. The \emph{Uniform} method denotes linear interpolation with fixed uniform weights.}%
  \label{tab:bwb-results}%
  \centering%
  \renewcommand{\arraystretch}{1.1}%
  \resizebox{\columnwidth}{!}{%
  \begin{tabular}{@{}lcccc@{}}%
    \toprule
                                 & Dynamic  && \multicolumn{2}{c}{Static} \\ \cmidrule{2-2} \cmidrule{4-5}
    Interpolation Method         & Val. PPL && Val. PPL & Test PPL\\ \midrule
    Uniform                      & -        && 99.1    & 97.5 \\
    Linear Interpolation         & 94.2     && 93.7    & 91.9 \\
    Count Merging                & 90.8     && 85.9    & 83.9 \\
    Bayesian Interpolation       & 87.4     && 85.1    & 83.2 \\
    \bottomrule
  \end{tabular}}
\end{table}

\begin{table}[th]
  \caption{Results on the WikiText scenario. The \emph{Uniform} method denotes linear interpolation with fixed uniform weights.}%
  \label{tab:wiki-results}%
  \centering%
  \renewcommand{\arraystretch}{1.1}%
  \resizebox{\columnwidth}{!}{%
  \begin{tabular}{@{}lcccc@{}}%
    \toprule
                                 & Dynamic  && \multicolumn{2}{c}{Static} \\ \cmidrule{2-2} \cmidrule{4-5}
    Interpolation Method         & Val. PPL && Val. PPL & Test PPL\\ \midrule
    Uniform                      & -        && 253.8    & 264.0 \\
    Linear Interpolation         & 228.1    && 226.1    & 235.6 \\
    Count Merging                & 254.8    && 221.6    & 232.0 \\
    Bayesian Interpolation       & 230.3    && 219.3    & 229.5 \\
    \bottomrule
  \end{tabular}}
\end{table}

\begin{table*}[!t]
  \caption{Results in the Smart Speaker scenario. The \emph{Uniform} method denotes linear interpolation with fixed uniform weights.}%
  \label{tab:ss-results}%
  \centering%
  \renewcommand{\arraystretch}{1.1}%
  \begin{tabular}{@{}lccccccccc@{}}
    \toprule
                                 & Dynamic  && \multicolumn{3}{c}{Static}        && \multicolumn{3}{c}{Pruned}\\ \cmidrule{2-2} \cmidrule{4-6} \cmidrule{8-10}
    Interpolation Method         &           Val. PPL && \#$n$-grams &           Val. PPL &           Test PPL && \#$n$-grams &           Test PPL & Test WER \\ \midrule
    Uniform                      &           -        && 634M        &           19.0     &           19.0     && 5.4M        &           21.5     & 5.3 \\
    Linear Interpolation         &           10.7     && 634M        &           10.7     &           10.7     && 5.9M        &           11.2     & 4.2 \\
    Count Merging                & \phantom{0}8.8     && 634M        & \phantom{0}8.8     & \phantom{0}8.8     && 7.7M        & \phantom{0}9.2     & 3.8 \\
    Bayesian Interpolation       & \phantom{0}8.8     && 634M        & \phantom{0}8.8     & \phantom{0}8.8     && 7.6M        & \phantom{0}9.2     & 3.8 \\
    \bottomrule
  \end{tabular}
\end{table*}
\subsection{Data}
We evaluated in three different scenarios.  We constructed the first scenario from the Billion Word dataset \cite{Chelba2013onebillionwords} and the second from the WikiText-103 dataset \cite{Merity2016}.  For the third scenario, we collected training data from a virtual assistant hand held device use case and targeted a virtual assistant smart speaker use case.

\subsubsection{Billion Word}
The Billion Word dataset consists of a single training set and a matched heldout set.  However, in order to perform informative experiments, we wanted a scenario with clusters of training data, each with unknown relevance to matched validation and test sets.  We did the following to create this kind of scenario from the Billion Word data.

First, we designated the first 10 partitions of the heldout set as test data and the remaining partitions as validation data.  Next, we jointly clustered the training, validation and test data.  To do this, we computed TF-IDF vectors on the first 20 partitions of the training data, then used Latent Semantic Analysis (LSA) to reduce the dimensionality of those vectors to 20 \cite{Deerwester}.  We then ran K-means to learn 10 cluster centroids \cite{MacQueen}.  After that, we assigned the remaining training, validation and test data to clusters using the IDF statistics, dimensionality reduction and cluster centroids learned on the 20 training data partitions.

To create a synthetic target use case, we then chose a random set of weights for the clusters from a Dirichlet distribution with $\alpha_0 = \alpha_1 = ... = \alpha_9 = 1$.  We created the test and validation sets by sampling respectively from the clustered test and validation data according to the randomly chosen weights.  

This process resulted in about 800M words of training data, 840K words of validation data and 215K words of test data.  We chose the vocabulary in the conventional way for this corpus, using all words that appeared in the training data three or more times.  This resulted in a vocabulary of 793K words.

\subsubsection{WikiText}
The WikiText-103 dataset \cite{Merity2016} consists of the normalized text of 28350 Wikipedia articles containing 101M words\footnote{Since we identified and removed exact duplicates of articles in the training set, our statistics are slightly different from those presented on the official webpage https://blog.einstein.ai/the-wikitext-long-term-dependency-language-modeling-dataset/}.  As in the Billion Word dataset, the WikiText-103 training set is monolithic.

For this corpus, we were able to use available metadata to create clusters.  Using the full vocabulary version of WikiText (called ``raw character level data''), we matched the page titles with the page titles from an English Wikipedia dump. We allowed for a fuzzy character match, ignoring certain issues of spacing, casing and punctuation, always choosing the most exact possible match.

Given the matched Wikipedia articles, we performed a depth-first search following the links to Wikipedia category pages, until we hit one of the Wikipedia main categories\footnote{https://en.wikipedia.org/wiki/Category:Main\_topic\_classifications} (28 broad top-down categories like "Nature" or "History").
We limited the maximum depth of the search and in case several main categories were found in the same depth, we chose the category with the fewest articles, as determined in a preliminary run.
The resulting clustering did not split the data evenly - the biggest clusters were "People" and "Entertainment", the smallest "Concepts" and "Humanities".

We limited the vocabulary to all words occurring three or more times in the training data, resulting in 266K words.  We randomly selected from the ``People'' cluster from the training data to create our test and validation sets and excluded this cluster from the training set we used in our experiments.  This process resulted in about 85M words of training data, 90K words of validation data and 90K words of test data.

\subsubsection{Smart Speaker}
In this scenario, the training data consisted of speech recognized transcripts from a virtual assistant handheld device use case that had been automatically clustered by a machine-learned classifier into 32 domains.  Examples of these domains are ``media player'', ``home automation'' and ``arithmetic.''  The validation and test sets were randomly sampled from a virtual assistant smart speaker use case and hand-transcribed.  The training data contained 57B words in total.  The validation and test data were comprised of 502K words and 434K words, respectively.  The vocabulary size was 570K.

\subsection{Setup}
In all experiments, we built 4-gram component LMs using Good-Turing smoothing \cite{Church}.  In the Smart Speaker experiments, minimum count thresholds of 2, 3 and 5 were applied for bigrams, trigrams and 4-grams respectively.  No count thresholds were applied in the other scenarios.  We optimized interpolation parameters using L-BFGS-B \cite{Byrd} to minimize validation set perplexity.

For speech recognition experiments, we used statically interpolated LMs, pruned using entropy pruning \cite{Stolcke}.  The acoustic model was a convolutional neural network with about 28M parameters, trained from millions of manually transcribed, anonymized virtual assistant requests. The input to the model was composed of 40 mel-spaced filter bank outputs; each frame was concatenated with the preceding and succeeding ten frames, giving an input of 840 dimensions.

In all experiments, we report validation set perplexities on the dynamically interpolated models and validation and test set perplexities on the unpruned statically interpolated models.  In speech recognition experiments, we also report test set perplexities on the pruned statically interpolated models, the number of $n$-grams in the unpruned and pruned models, and test set word error rates (WERs) on the pruned models.  As a point of reference, we include results for a uniform linear interpolation.

\subsection{Results}
Tables \ref{tab:bwb-results} and \ref{tab:wiki-results} show perplexity results for dynamically and statically interpolated models for the Billion Word and WikiText scenarios.  Focusing on the statically interpolated models, we see that the differences between interpolation method perplexities on the validation and test sets are roughly similar.  We observe that count merging and Bayesian interpolation outperform linear interpolation, as expected.  The difference in performance is larger in the Billion Word scenario, perhaps because the validation and test data are better matched to the training sources.  Count merging and Bayesian interpolation perform comparably.

We notice that in the WikiText scenario the test set perplexities are higher than validation set perplexities.  This is likely a consequence of characteristics of the WikiText data and how the validation and test sets were selected.  The validation and test set data were created by randomly sampling lines from a single data cluster.  A single line often contains many words (e.g. an entire paragraph.)  Thus, even with 90K words in each, the test and validation sets are not perfectly matched.

Looking at the perplexities of the dynamically interpolated models, we notice higher perplexities for count merging and Bayesian interpolation when comparing to static interpolation.  The difference is especially striking for count merging in the WikiText scenario.  In the case of count merging, we believe this is explained by the following phenomenon.  Consider an $n$-gram with these properties:
\begin{enumerate}
\item \label{it:lp} The component models corresponding to the data sets with non-zero history counts assign the $n$-gram a very low probability (e.g. because the predicted word is unseen.)
\item \label{it:nc} The $n$-gram is not in any of the component models.
\end{enumerate}
The dynamically interpolated model will assign the $n$-gram a very low probability, because of \ref{it:lp}.  However, this $n$-gram will not be included in the statically interpolated model, because of \ref{it:nc}.  Thus, the statically interpolated model will obtain the probability for this $n$-gram from a lower order $n$-gram (combined with the backoff probability for the history.)  The lower-order $n$-gram will have been present in at least one of the component models, and so the probability assigned by the statically interpolated model is likely to be larger than the one assigned by the dynamically interpolated model.

An analogous situation arises for Bayesian interpolation, but the effect is much less severe.  This is a positive consequence of the smooth estimate of $p(h|i)$ used by Bayesian interpolation.  While $p^{\text{BI}}(h|i)$ may be small for a given $i$, it will never be zero.

While we do not address the issue here, this analysis suggests modifying count merging and Bayesian interpolation to take into account the $n$-gram structure of the final interpolated model.  More specifically, the optimization could compute the probability for a particular validation set $n$-gram based on the highest order $n$-gram present in any of the component models.

Table \ref{tab:ss-results} shows perplexity and WER results for the Smart Speaker scenario.  We see that perplexity results on the unpruned static models are consistent with the results observed in the Billion Word and WikiText scenarios.  Further, the perplexity improvements on the statically interpolated models carry over after pruning.  Lastly, test set WER results are consistent with the perplexity results.  We see a 9.5\% relative WER reduction between linear interpolation and count merging or Bayesian interpolation and no significant difference between count merging and Bayesian interpolation.

\subsection{Other Considerations}
There are considerations beyond performance when comparing count merging and Bayesian interpolation.  Count merging requires storing history counts for each component LM, while Bayesian interpolation does not.  On the other hand, computing the history statistic is more computationally expensive for Bayesian interpolation than count merging.  In Bayesian interpolation, multiple $n$-gram probabilities must be retrieved, while in the case of count merging, only a single count must be retrieved.  This may introduce significant additional computation for Bayesian merging when creating a statically interpolated LM from very large component LMs.  We suspect that in most circumstances the added complexity of storing history counts in count merging will be the bigger concern.

It is also useful to consider how the two interpolation techniques might perform in less abundant data scenarios or cases where the validation set is poorly matched to the training data clusters.  In those conditions, $c_i(h)$ will be $0$ for many of the word histories in the validation set, making $p^{\text{CM}}(h|i)$ a very poor estimate of $p(h|i)$.  One can think of heuristics to mitigate this problem, but Bayesian interpolation will handle these conditions gracefully without modification.

\section{Conclusions}
\label{sec:conclusions}
We showed a theoretical connection between count merging and Bayesian interpolation.  We evaluated these techniques as well as linear interpolation in three scenarios with abundant training data.  Consistent with prior work, our results indicate that both count merging and Bayesian interpolation outperform linear interpolation.  Count merging and Bayesian interpolation perform comparably, but Bayesian interpolation has other advantages that will make it the preferred approach in most circumstances.


\bibliographystyle{IEEEtran}
\bibliography{mybib}

\begin{thebibliography}{10}
\providecommand{\url}[1]{#1}
\csname url@samestyle\endcsname
\providecommand{\newblock}{\relax}
\providecommand{\bibinfo}[2]{#2}
\providecommand{\BIBentrySTDinterwordspacing}{\spaceskip=0pt\relax}
\providecommand{\BIBentryALTinterwordstretchfactor}{4}
\providecommand{\BIBentryALTinterwordspacing}{\spaceskip=\fontdimen2\font plus
\BIBentryALTinterwordstretchfactor\fontdimen3\font minus
  \fontdimen4\font\relax}
\providecommand{\BIBforeignlanguage}[2]{{%
\expandafter\ifx\csname l@#1\endcsname\relax
\typeout{** WARNING: IEEEtran.bst: No hyphenation pattern has been}%
\typeout{** loaded for the language `#1'. Using the pattern for}%
\typeout{** the default language instead.}%
\else
\language=\csname l@#1\endcsname
\fi
#2}}
\providecommand{\BIBdecl}{\relax}
\BIBdecl

\bibitem{Chelba2013onebillionwords}
C.~Chelba, T.~Mikolov, M.~Schuster, Q.~Ge, T.~Brants, P.~Koehn, and
  T.~Robinson, ``One billion word benchmark for measuring progress in
  statistical language modeling,'' in \emph{Proc. {INTERSPEECH}}, 2014.

\bibitem{JelinekMercer80}
F.~Jelinek and R.~Mercer, ``Interpolated estimation of {Markov} source
  parameters from sparse data,'' \emph{Proc. Workshop Pattern Recognition in
  Practice}, pp. 381--397, May 1980.

\bibitem{Bacchiani}
M.~Bacchiani, M.~Riley, B.~Roark, and R.~Sproat, ``{MAP} adaptation of
  stochastic grammars,'' \emph{Computer speech \& language}, vol.~20, no.~1,
  pp. 41--68, 2006.

\bibitem{Hsu}
B.-J. Hsu, ``Generalized linear interpolation of language models,'' in
  \emph{IEEE Workshop on Automatic Speech Recognition \& Understanding}, 2007.

\bibitem{Allauzen}
C.~Allauzen and M.~Riley, ``Bayesian language model interpolation for mobile
  speech input,'' in \emph{Proc. {INTERSPEECH}}, 2011.

\bibitem{Bacchiani2003}
M.~Bacchiani and B.~Roark, ``Unsupervised language model adaptation,'' in
  \emph{Proc. of {ICASSP}}, 2003.

\bibitem{Weintraub}
M.~Weintraub, Y.~Aksu, S.~Dharanipragada, S.~Khudanpur, H.~Ney, J.~Prange,
  A.~Stolcke, F.~Jelinek, and E.~Shriberg, ``{LM95} project report: Fast
  training and portability,'' Research Note 1, Center for Language and Speech
  Processing, Johns Hopkins University, Tech. Rep., 1996.

\bibitem{Liu}
X.~Liu, M.~Gales, J.~Francis, and P.~C. Woodland, ``Use of contexts in language
  model interpolation and adaptation,'' \emph{Computer Speech \& Language},
  vol.~27, no.~1, pp. 301--321, 2013.

\bibitem{Merity2016}
S.~{Merity}, C.~{Xiong}, J.~{Bradbury}, and R.~{Socher}, ``{Pointer sentinel
  mixture models},'' in \emph{Proc. {ICLR}}, 2017.

\bibitem{Deerwester}
S.~Deerwester, S.~T. Dumais, G.~W. Furnas, T.~K. Landauer, and R.~Harshman,
  ``Indexing by latent semantic analysis,'' \emph{Journal of the American
  society for information science}, vol.~41, no.~6, pp. 391--407, 1990.

\bibitem{MacQueen}
J.~MacQueen \emph{et~al.}, ``Some methods for classification and analysis of
  multivariate observations,'' in \emph{Proceedings of the fifth Berkeley
  symposium on mathematical statistics and probability}, vol.~1, no.~14.\hskip
  1em plus 0.5em minus 0.4em\relax Oakland, CA, USA, 1967, pp. 281--297.

\bibitem{Church}
K.~W. Church and W.~A. Gale, ``A comparison of the enhanced {Good-Turing} and
  deleted estimation methods for estimating probabilities of {English}
  bigrams,'' \emph{Computer Speech \& Language}, vol.~5, no.~1, pp. 19--54,
  1991.

\bibitem{Byrd}
R.~H. Byrd, P.~Lu, J.~Nocedal, and C.~Zhu, ``A limited memory algorithm for
  bound constrained optimization,'' \emph{SIAM Journal on Scientific
  Computing}, vol.~16, no.~5, pp. 1190--1208, 1995.

\bibitem{Stolcke}
A.~Stolcke, ``Entropy-based pruning of backoff language models,'' in
  \emph{Proc. of DARPA Broadcast News Transcription and Understanding
  Workshop}, 1998.

\end{thebibliography}

\end{document}